\documentclass[journal]{IEEEtran}
\usepackage[utf8]{inputenc}
\usepackage{graphicx}
\usepackage{amsmath}
\usepackage{amssymb}
\usepackage{tikz}
\usetikzlibrary{shapes.geometric, arrows.meta, positioning, calc, fit, backgrounds}
\usepackage{circuitikz}
\usepackage{pgfplots}
\pgfplotsset{compat=1.18}
\usepackage{booktabs}
\usepackage{tabularx}
\usepackage{siunitx}
\usepackage[autostyle]{csquotes}
\MakeOuterQuote{"}
\usepackage[style=numeric, backend=biber, sorting=none]{biblatex}
\usepackage[hidelinks]{hyperref}

\begin{document}

\title{Estimating Available Traction Power in Multi-Train AC Railway Networks from a Distance-Dependent Power Envelope}

\author{
	\IEEEauthorblockN{Márton László Ambrus, Xiao Liu, Stuart Hillmansen, and Zhongbei Tian}
	\thanks{The authors are with the Birmingham Centre for Railway Research and Education (BCRRE), Department of Electronic, Electrical and Systems Engineering, University of Birmingham, Birmingham, U.K. (e-mail: m.l.ambrus@bham.ac.uk).}
}

\maketitle

\begin{abstract}
Decarbonisation is raising the electrical load on mainline alternating-current railway feeders that were not designed for sustained, simultaneous high-power demand. When several trains accelerate together on a shared feeder, the contact-line voltage can fall far enough to trigger rolling-stock current limitation or feeder protection, eroding capacity and reliability. Preventing this in real time requires a quantity conventional operation does not expose: a localised, continuously updated estimate of the traction power available to each train given the live network state. A railway power-flow model, with trains represented under a voltage-dependent automatic current-limitation characteristic, shows that the minimum network voltage is governed by the product of power and distance rather than by power alone, yielding a distance-dependent single-train power envelope. This envelope does not add up when several trains share a feeder, so a conservative pairwise screen is generalised to a solver-free multi-train estimate: a calibrated shared-path voltage model returning the minimum section voltage and the per-train available power for any number of trains. Calibration uses two short offline solver runs, one fixing the self-impedance and one the inter-train coupling through a separation-dependent factor. Its current-limitation behaviour follows EN 50388-1, and on matched multi-train cases the estimate tracks the full power flow to within about nine per cent on average across two-, three-, and four-train cases, improving as more trains share the feeder, while its online cost scales with the number of trains rather than the network size.
\end{abstract}


\section{Introduction}\label{sec:intro}
\IEEEPARstart{T}{he} decarbonisation of surface transport places rail in a central but demanding position. Electrified rail is already among the most efficient ways to move people and freight, and national strategy assumes that electric traction, rather than reduced service, will deliver the required emissions savings. The corollary is that operators must extract more capacity from electrical infrastructure that was, in many cases, dimensioned for lighter and less concurrent loading than today's mixed-traffic timetables impose.

On mixed-traffic AC corridors the binding constraint is increasingly electrical rather than signalling-related. A train accelerating at high power draws a large current that, acting through the feeder impedance, depresses the catenary voltage over a wide section of route. When a second high-power movement overlaps the first, the combined demand can drag the pantograph voltage low enough to invoke the rolling stock's protective current limitation or, in the worst case, to operate feeder protection and disconnect a train. Such events waste capacity, degrade reliability, and are difficult to anticipate from the timetable alone because they depend jointly on where trains are and how much power they draw at the same instant.

Mitigating these events in real time requires information that conventional operation does not make explicit: at any given moment, how much additional traction power can a given train draw without pushing the network past its voltage or loading limits? A comprehensive survey of algorithmic energy management for constrained railway networks \cite{ambrus2026review} shows that, although multi-train, grid-coupled optimisation is well studied offline, there is no established method that exposes this \emph{available traction power} as a quantity computable quickly enough, and from data operationally available, to guide live operation. Existing grid-coupled formulations recover the full electrical state by solving a power flow, which is too slow for the control loop; single-train methods are fast but ignore the inter-train coupling that causes the problem.

This paper addresses that gap by making available traction power observable in real time. Framed in transportation terms, the method supplies a continuously updated estimate of the electrical state, the power headroom at each train, fast enough to serve as an input to intelligent control, delivered through the same connected driver advisory channel that already carries speed and timing advice, rather than a quantity recoverable only by offline network solution. Its contributions are:
\begin{enumerate}
	\item an AC power-flow model of a mixed-traffic traction feeder in which trains are represented under the voltage-dependent automatic current-limitation (ACL) characteristic of EN 50388-1 \cite{en50388}, reproducing the low-voltage behaviour that matters for capacity;
	\item identification of the power-$\times$-distance coupling that governs minimum network voltage, which shows why a uniform power limit is ineffective and a location-aware limit is required;
	\item a rigorous single-train available-power envelope $P_{\max}(d)$, together with the finding that it does not extend additively to multiple trains and the electrical explanation of why, and a conservative two-train screen retained for the dominant interaction; and
	\item verification of the model's low-voltage current-limitation behaviour against the EN 50388-1 characteristic, and a demonstration of the envelope and headroom estimates on a representative corridor, including their computational cost relative to a full online power flow.
\end{enumerate}

The remainder of the paper is organised as follows. Section \ref{sec:related} positions the work. Section \ref{sec:model} sets out the network and train models. Section \ref{sec:analysis} establishes the power-$\times$-distance coupling. Section \ref{sec:method} derives the power envelope and headroom estimate. Sections \ref{sec:setup} and \ref{sec:results} describe the representative corridor and the verification and demonstration results. Sections \ref{sec:discussion} and \ref{sec:conclusion} discuss limitations and conclude.

\section{Related Work}\label{sec:related}
Four strands of prior work bear on real-time available-power estimation: energy-efficient train operation, train-power co-simulation, AC traction power-flow modelling, and the acceleration of power-flow and optimal-power-flow computation. A broader survey of algorithmic energy management in constrained railway networks is given in \cite{ambrus2026review}; the present section is confined to the modelling and estimation foundations on which this paper builds.

\emph{Energy-efficient driving and timetabling.} The largest body of work optimises train speed profiles and timetables for energy. Single-train trajectory optimisation by dynamic programming and related methods is mature \cite{6410425,SCHEEPMAKER2017355} and has been extended to multiple trains sharing regenerated energy \cite{7039269,6856141}, to combined speed-profile and timetable optimisation \cite{ran2020energy,9006917}, and to data-driven and reinforcement-learning controllers \cite{8649821,shang2022energy,peng2024energy,liu2025comparative}. System-level reviews \cite{SCHEEPMAKER2017355,DOUGLAS20151149} and energy-storage measures \cite{ratniyomchai2014energy,7327077} complete the mitigation toolkit, and connected driver advisory systems provide the channel through which such advice is delivered in service \cite{cdas_das}. This work is predominantly offline: it computes optimal references ahead of operation and, where electrical limits enter at all, treats them as fixed constraints rather than as a live, location-dependent quantity.

\emph{Train-power simulation.} Coupling a multi-train movement model to a power-supply network solved by AC power flow is well established, from the early rapid-transit simulators \cite{mellitt1978simulator,goodman2008overview} to modern co-simulation used for energy and capacity studies \cite{tian2017system,zhao_2013_railway,en13112788}. Movement-model calibration is itself a studied problem \cite{cunillera2023literature}, and train-following and virtual-coupling models capture the longitudinal interaction between services \cite{quaglietta2020multi}. These tools reproduce the dependence of achievable power on pantograph voltage faithfully, but are used for offline analysis rather than as online estimators.

\emph{AC traction power supply and power flow.} The \SI{25}{\kilo\volt} AC supply and its $2\times\SI{25}{\kilo\volt}$ autotransformer variant are described in \cite{bhargava1999railway,white2015ac,oura1998railway,tao_2024_comprehensive}, with EMC and design considerations in \cite{fei2020ac,11364793}. Dedicated railway power-flow formulations exploit the chain topology of the feeder \cite{powerflow_railway} and use current-injection schemes for traction networks \cite{mohamed2017modified,9006917}. The low-voltage behaviour of rolling stock is standardised through the automatic current-limitation characteristic of EN 50388-1 \cite{en50388}, in conjunction with the supply-voltage limits of EN 50163 \cite{en50163}; the resulting power limitations are examined in related industry work \cite{rssb_t1331}. These solvers recover the full electrical state accurately, but at a cost set by the size of the network.

\emph{Fast power-flow and OPF evaluation.} Because repeated power-flow or optimal-power-flow solution is the bottleneck for online use, a long line of work has sought faster approximations, from classical linearised and outage-screening methods \cite{peterson1972iterative,dommel2007optimal,bukhsh2013local} to recent learning-based AC-OPF surrogates \cite{zamzam2020learning,zhou2020data}. These accelerate the \emph{general} power-system problem; none returns the railway-specific quantity of interest here, the location-aware traction power available to each train.

Across these strands the fast methods are single-train or neglect inter-train coupling, while the methods that capture the coupling require a network solve. What is missing, and what this paper supplies, is a means of expressing the power available to each train as a closed-form, location-aware quantity that can be evaluated within the control loop rather than by repeated solution of the network power flow.

\section{Traction Power Network Model}\label{sec:model}

\subsection{Feeder Architecture}
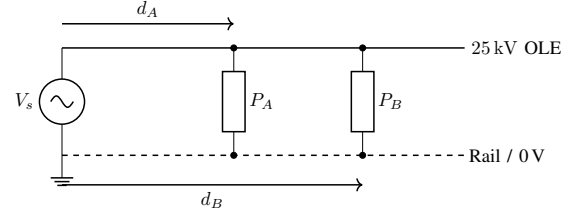
\begin{figure}[!t]
	\centering
	\resizebox{0.85\columnwidth}{!}{
		\begin{circuitikz}
			\draw[thick] (0,2) -- (7.5,2) node[right] {\SI{25}{\kilo\volt} OLE};
			\draw[thick, dashed] (0,0) -- (7.5,0) node[right, fill=white, inner sep=2pt] {Rail / 0\,V};
			\draw (0,0) to[sV, l=$V_s$] (0,2);
			\draw (0,0) -- (-0.0,0) node[ground]{};
			\draw (3.2,2) to[generic, l=$P_A$, *-*] (3.2,0);
			\draw (5.6,2) to[generic, l=$P_B$, *-*] (5.6,0);
			\draw[->, thick] (0,2.45) -- (3.2,2.45) node[midway, above]{$d_A$};
			\draw[->, thick] (0,-0.55) -- (5.6,-0.55) node[midway, below]{$d_B$};
		\end{circuitikz}
	}
	\caption{Representative radial \SI{25}{\kilo\volt} feeder with two trains at distances $d_A$ and $d_B$ from the supply point. Series impedance per unit length is $z=r+\mathrm{j}x$.}
	\label{fig:feeder}
\end{figure}

A radial feeding arrangement is considered, in which a single \SI{25}{\kilo\volt} AC feeder station supplies a length of double-track route, with trains distributed along the feeder at distances $d$ from the supply point (Fig. \ref{fig:feeder}). The treatment extends directly to autotransformer ($2\times\SI{25}{\kilo\volt}$) feeding by replacing the single series impedance with the equivalent catenary-feeder network; the radial case is retained here because it exposes the distance dependence most clearly. The feeder is characterised by a series impedance per unit length $z = r + \mathrm{j}x$, and the supply point is modelled as a stiff source of voltage $V_{s}$ behind the source impedance of the feeder transformer.

\subsection{Multi-Train Movement Model}
Each train is advanced by a longitudinal motion model. With $v$ the speed, $m$ the mass, $\lambda$ the rotational-mass allowance, and $i_{\mathrm g}$ the gradient in per mille, Newton's second law gives
\begin{equation}
	(1+\lambda)\,m\,\frac{\mathrm{d}v}{\mathrm{d}t} = F_t(v) - R(v) - m g \frac{i_{\mathrm g}}{1000},
	\label{eq:motion}
\end{equation}
where the running resistance follows the Davis form
\begin{equation}
	R(v) = A + B v + C v^{2},
	\label{eq:davis}
\end{equation}
and the tractive effort is bounded by the adhesion limit and the rated power,
\begin{equation}
	F_t(v) = \min\!\left( F_{\max},\; \frac{\eta_t P_{\mathrm{r}}}{v} \right).
	\label{eq:te}
\end{equation}
The electrical power demanded at the pantograph during motoring is
\begin{equation}
	P_{\mathrm{dem}} = \frac{F_t(v)\,v}{\eta_t} + P_{\mathrm{aux}},
	\label{eq:pdem}
\end{equation}
with $\eta_t$ the traction-chain efficiency and $P_{\mathrm{aux}}$ the auxiliary load; regenerative braking is represented by a negative demand subject to the receptivity of the network.

\subsection{Voltage-Dependent Automatic Current Limitation}
The behaviour that distinguishes a credible low-voltage model from a constant-power approximation is the rolling stock's automatic current limitation. Following EN 50388-1, a train does not draw unlimited current as the voltage falls; instead its permissible current is capped by a characteristic $I_{\lim}(V)$ that derates from the rated value as the pantograph voltage drops between an upper and a lower threshold,
\begin{equation}
	I_{\lim}(V) =
	\begin{cases}
		I_{\mathrm{r}}, & V \ge V_1,\\[2pt]
		I_{\mathrm{r}}\dfrac{V - V_0}{V_1 - V_0}, & V_0 \le V < V_1,\\[6pt]
		0, & V < V_0,
	\end{cases}
	\label{eq:acl}
\end{equation}
where the upper threshold at which derating begins is $V_1 \approx \SI{19}{\kilo\volt}$ and the lower threshold below which current is fully suppressed is $V_0 \approx \SI{12.5}{\kilo\volt}$; the EN 50163 lowest non-permanent voltage, $U_{\min}=\SI{17.5}{\kilo\volt}$, lies between the two and is retained as the voltage-compliance limit against which available power is assessed (Section \ref{sec:method}), so that a train may continue to draw a reduced current in the non-compliant band $V_0 \le V < U_{\min}$ rather than disconnecting at $U_{\min}$. The current actually drawn is then $I_k = \min\!\big(I_{\mathrm{dem},k},\, I_{\lim}(V_k)\big)$, so that under depressed voltage the delivered power falls below the demand of \eqref{eq:pdem}. A constant-power baseline, in which \eqref{eq:acl} is not enforced, is retained only for reproducing legacy protection-coordination events; it is not representative of EN 50388-1 rolling stock and is identified as such where used.

\subsection{AC Power-Flow Solution}
At each simulation time step the trains present a set of complex power injections $S_k = P_k + \mathrm{j}Q_k$ at their respective nodes, the reactive component following from the assumed displacement factor. Collecting the nodal voltages in $\mathbf{V}$ and forming the network admittance matrix $\mathbf{Y}$, the network satisfies
\begin{equation}
	\mathbf{Y}\mathbf{V} = \mathbf{I}, \qquad I_k = \left( \frac{S_k}{V_k} \right)^{\!*},
	\label{eq:pf}
\end{equation}
which is nonlinear because the injected currents depend on the unknown voltages. It is solved iteratively by a current-injection scheme: from the voltages of iteration $n-1$ the injections are evaluated, the linear system is solved for $\mathbf{V}^{(n)}$, the limitation \eqref{eq:acl} is applied, and the process repeats until $\lVert \mathbf{V}^{(n)} - \mathbf{V}^{(n-1)} \rVert < \varepsilon$. Convergence is rapid under normal loading but degrades sharply when several trains draw high power at low voltage, a numerical signature of the same physical stress that the method seeks to avoid (Section \ref{sec:results}).

\section{Voltage Behaviour Under Mixed Traffic}\label{sec:analysis}
The motivation for a location-aware power limit follows directly from the model. For a single train at distance $d$ drawing active power $P$ at pantograph voltage $V$, the resistive voltage drop along the feeder is approximately
\begin{equation}
	V_{s} - V \;\approx\; r\,d\,\frac{P}{V},
	\label{eq:drop}
\end{equation}
so that the voltage depression scales not with power alone but with the product of power and distance. Two consequences matter for operation. First, a train far from the supply point depresses the voltage far more, for the same power, than one nearby. Second, when the high-power windows of two trains overlap, their drops superpose, and the worst voltage occurs when high demand coincides with large distance.

This explains an effect that is otherwise counterintuitive: uniformly reducing acceleration does not monotonically improve the minimum voltage. Lowering a train's acceleration reduces its peak power but also lengthens its high-power window and shifts it in time and position; a moderate reduction can therefore relocate the worst-case overlap to a less favourable point on the feeder, leaving the minimum voltage no better, or worse, than at full acceleration. Only a reduction large enough to lower peak power outright improves the minimum voltage, and applying such a reduction to every train indiscriminately incurs unacceptable journey-time penalties. The implication is that an effective limit must be applied selectively, to the trains and locations that actually bind, which requires the power available at each location to be known. That quantity is derived next.

\section{Online Available Traction Power Estimation}\label{sec:method}

\subsection{Distance-Dependent Power Envelope}
\begin{figure}[!t]
	\centering
	\includegraphics[width=\columnwidth]{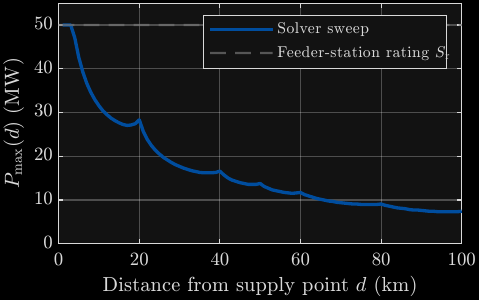}
	\caption{Distance-dependent power envelope $P_{\max}(d)$ for the corridor of Section \ref{sec:setup}, obtained by sweeping a single train's power demand to the voltage-compliance limit at each distance with the full power flow: rating-limited at $S_{\mathrm r}=\SI{50}{\mega\watt}$ near the supply point, then voltage-limited and decaying with distance.}
	\label{fig:envelope}
\end{figure}

Define $P_{\max}(d)$ as the greatest active power that can be delivered to a single train at distance $d$ without violating either the supply-point rating or the lower voltage-compliance limit $U_{\min}$. Near the supply point the binding constraint is the feeder-station rating $S_{\mathrm{r}}$; further out it is the voltage limit. Imposing $V = U_{\min}$ in \eqref{eq:drop} and solving for power gives the voltage-limited branch, so that
\begin{equation}
	P_{\max}(d) = \min\!\left( S_{\mathrm{r}},\; \frac{U_{\min}\,(V_{s} - U_{\min})}{r\,d} \right).
	\label{eq:envelope}
\end{equation}
Equation \eqref{eq:envelope} is the closed-form insight; in practice $P_{\max}(d)$ is evaluated numerically from the power-flow model of Section \ref{sec:model}, which captures reactive flow, the autotransformer network where present, and the current-limitation characteristic. The result is a curve that is flat at $S_{\mathrm{r}}$ near the supply point and decays with distance thereafter (Fig. \ref{fig:envelope}). Because it depends only on fixed infrastructure parameters, it is computed once, offline, and stored as a lookup table.

\subsection{From Envelope to a Conservative Headroom Screen}
The envelope $P_{\max}(d)$ is exact for a single train, and it is tempting to extend it to $N$ trains by treating their demands as additive, giving a candidate per-train headroom
\begin{equation}
	\tilde{H}_i = P_{\max}(d_i) - \sum_{k \neq i} P_k .
	\label{eq:headroom_naive}
\end{equation}
This is electrically invalid for more than one train. On a shared radial feeder the current in each segment is the sum of the currents of all trains beyond it, so the voltage at any train depends on the \emph{positions} of the other trains as well as their power; as Section \ref{sec:analysis} showed, it is the product of power and distance that governs the voltage, not power alone. A given amount of power drawn close to the supply point and the same power drawn far along the feeder therefore load the limiting segment unequally, and subtracting undifferentiated power from a single-train envelope, as in \eqref{eq:headroom_naive}, conflates the two. Direct evaluation across three or more trains confirms that \eqref{eq:headroom_naive} mis-estimates the feasible power, and it is not adopted.

What is retained is a conservative two-train screen. With a leading reference train A at distance $d_A$ drawing $P(d_A)$, the spare power at that location is
\begin{equation}
	H = P_{\max}(d_A) - P(d_A),
	\label{eq:headroom}
\end{equation}
which caps the command of a following train so that the dominant pairwise interaction cannot drive the reference location beyond its envelope. The screen deliberately addresses the worst single interaction rather than the full $N$-train state; it is adequate for flagging the binding events but is not a general multi-train headroom. It is retained not as a rival to the estimate developed next but as its cheapest special case: a calibration-free, single-lookup bound, one envelope evaluation and one subtraction, with no impedance fit and no iteration. Well suited to a transparent conservative guardrail and to a per-pair constraint in the demand-shaping optimisation, where a simple linear bound is easier to embed than an implicit solve. The general $N$-train estimate, more accurate but requiring the impedance calibration and a short iteration, is developed in Section \ref{sec:multitrain}, where the single-train envelope is generalised to a calibrated shared-path model evaluated without a power-flow solver.

\subsection{Solver-Free Multi-Train Estimation by a Calibrated Shared-Path Model}\label{sec:multitrain}
The non-additivity of the envelope does not, however, force a full online power flow. Because the voltage depression is governed by the product of power and the \emph{shared} feeder length, the voltage at every train can be written in closed form. For $N$ trains at distances $d_i$ drawing complex power $S_k$,
\begin{equation}
	V_i = V_s - \sum_{k} M_{ik}\left(\frac{S_k}{V_k}\right)^{\!*}, \qquad S_k = P_k\,(1+\mathrm{j}\kappa),
	\label{eq:sharedpath}
\end{equation}
where $\kappa=\tan(\arccos\varphi)$ fixes the reactive component from the displacement factor $\varphi$, and $\mathbf{M}$ is the calibrated shared-path impedance matrix introduced below. Its diagonal $M_{ii}$ is the impedance the supply point presents to train $i$ acting alone; its off-diagonal $M_{ik}$ ($i\neq k$) is the mutual term that makes train $k$'s current depress train $i$'s voltage, built on $\min(d_i,d_k)$, the length of feeder the two trains share back to the supply point, through which both their currents flow. The coupling that invalidated \eqref{eq:headroom_naive} is thus retained: a train's voltage depends on \emph{where} the other trains are, not merely on how much they draw. The series impedance from the supply point out to distance $d$ is taken as
\begin{equation}
	Z(d) = Z_0 + \beta(d)\,z\,d,
	\label{eq:zeff}
\end{equation}
with $Z_0$ the supply-point (transformer) impedance, $z$ the single-track per-unit-length impedance of Section \ref{sec:model}, and $\beta(d)$ a dimensionless scaling that absorbs the effect of double-track feeding and paralleling posts. The matrix entries are then
\begin{equation}
	M_{ik} = \gamma(d_i,d_k)\,Z\!\big(\min(d_i,d_k)\big), \qquad \gamma(d_i,d_i)\equiv 1,
	\label{eq:Mmatrix}
\end{equation}
so the diagonal is $M_{ii}=Z(d_i)$ unchanged, and the off-diagonal carries a mutual coupling-reduction factor $\gamma\le 1$.

\paragraph*{Two-stage calibration}
The two terms are calibrated separately, each from a short offline run of the power-flow model, because a single-train sweep observes only the diagonal: with one train present, \eqref{eq:sharedpath} reduces to the single-train relation behind the envelope $P_{\max}(d)$, and no single-train experiment ever excites an off-diagonal entry. The \emph{first} stage therefore recovers $\beta(d)$: given the solver's envelope value at each distance, \eqref{eq:zeff} is inverted for the $\beta(d)$ that reproduces it. For the representative double-track corridor $\beta(d)\approx 1$ close to the supply point where, ahead of the first paralleling post, the line behaves as a single track and settles to roughly $0.4$ further out, where the parallel paths approximately halve the effective impedance. By construction the model then reproduces the single-train envelope of Fig. \ref{fig:envelope} exactly.

The \emph{second} stage fixes the off-diagonal, which the first cannot see. Were the network a clean radial feeder, the shared-path term would be exact with $\gamma\equiv 1$; but the paralleling posts and track bonding that reduce the \emph{self}-impedance (the $\beta<1$ above) also weaken the \emph{mutual} coupling between separated trains, by an amount that grows with their separation. Left uncorrected, $\gamma\equiv 1$ over-couples the trains: maximising one drives the others' voltages down almost as fast as its own, so a distant train binds prematurely and the estimate is pessimistic. We model the reduction as a separation-dependent factor
\begin{equation}
	\gamma(d_i,d_k) = \gamma_\infty + (1-\gamma_\infty)\,\exp\!\big(-|d_i-d_k|/L\big),
	\label{eq:gamma}
\end{equation}
which is unity for coincident trains (preserving the single-train calibration on the diagonal) and decays towards a floor $\gamma_\infty$ as the separation grows. The two parameters $(\gamma_\infty,L)$ are fitted once, offline, by matching the estimate's two-train available power to the solver's over a handful of two-train sweeps. A leading train held at a fixed location and power while a probe train is swept along the route, minimising the squared headroom error. This adds one short solver pass beyond the envelope sweep and leaves the diagonal, and hence the single-train envelope, untouched.

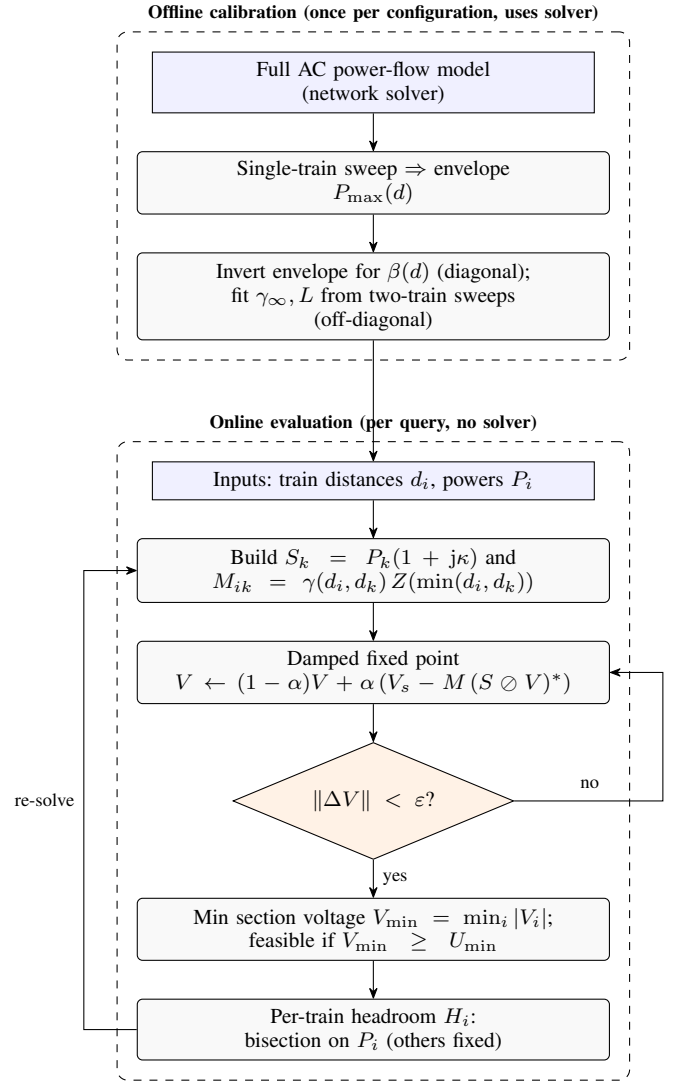
\begin{figure}[!t]
	\centering
	\begin{tikzpicture}[
		font=\footnotesize,
		>={Stealth[round]},
		node distance=5mm,
		proc/.style={rectangle, draw, rounded corners=2pt, align=center, text width=60mm, inner sep=3.5pt, fill=gray!5},
		io/.style={rectangle, draw, align=center, text width=56mm, inner sep=3.5pt, fill=blue!6},
		dec/.style={diamond, draw, aspect=2.4, align=center, inner sep=1pt, text width=28mm, fill=orange!10}
	]
		\node[io] (N1) {Full AC power-flow model\\ (network solver)};
		\node[proc, below=of N1] (N2) {Single-train sweep $\Rightarrow$ envelope\\ $P_{\max}(d)$};
		\node[proc, below=of N2] (N3) {Invert envelope for $\beta(d)$ (diagonal);\\ fit $\gamma_\infty,L$ from two-train sweeps\\ (off-diagonal)};
		\node[io, below=16mm of N3] (N4) {Inputs: train distances $d_i$, powers $P_i$};
		\node[proc, below=of N4] (N5) {Build $S_k=P_k(1+\mathrm{j}\kappa)$ and\\ $M_{ik}=\gamma(d_i,d_k)\,Z(\min(d_i,d_k))$};
		\node[proc, below=of N5] (N6) {Damped fixed point\\ $V\leftarrow(1-\alpha)V+\alpha\,(V_s-M\,(S\oslash V)^{*})$};
		\node[dec, below=of N6] (N7) {$\lVert\Delta V\rVert<\varepsilon$?};
		\node[proc, below=of N7] (N8) {Min section voltage $V_{\min}=\min_i|V_i|$;\\ feasible if $V_{\min}\ge U_{\min}$};
		\node[proc, below=of N8] (N9) {Per-train headroom $H_i$:\\ bisection on $P_i$ (others fixed)};
		\draw[->] (N1)--(N2); \draw[->] (N2)--(N3); \draw[->] (N3)--(N4);
		\draw[->] (N4)--(N5); \draw[->] (N5)--(N6); \draw[->] (N6)--(N7);
		\draw[->] (N7)-- node[right,font=\scriptsize]{yes} (N8);
		\draw[->] (N8)--(N9);
		\draw[->] (N7.east) -| ([xshift=7mm]N6.east) -- (N6.east);
		\node[font=\scriptsize] at ([xshift=10mm,yshift=2.2mm]N7.east) {no};
		\draw[->] (N9.west) -- ++(-7mm,0) |- node[left,pos=0.25,font=\scriptsize]{re-solve} (N5.west);
		\begin{scope}[on background layer]
			\node[fit=(N1)(N2)(N3), draw, dashed, rounded corners, inner sep=2.6mm, label={[font=\scriptsize\bfseries]above:Offline calibration (once per configuration, uses solver)}] {};
			\node[fit=(N4)(N5)(N6)(N7)(N8)(N9), draw, dashed, rounded corners, inner sep=2.6mm, label={[font=\scriptsize\bfseries]above:Online evaluation (per query, no solver)}] {};
		\end{scope}
	\end{tikzpicture}
	\caption{Functional flow of the solver-free headroom calculator. A single offline power-flow sweep yields the envelope $P_{\max}(d)$, from which the diagonal scaling $\beta(d)$ is recovered; a short set of two-train sweeps then fixes the off-diagonal coupling factor $\gamma(d_i,d_k)$. Thereafter each query builds the calibrated shared-path impedance matrix, solves a damped fixed point for the train voltages, and returns the minimum section voltage and per-train headroom with no further power-flow solution.}
	\label{fig:flow}
\end{figure}

Equation \eqref{eq:sharedpath} is implicit in $\mathbf{V}$ but is solved cheaply by the damped fixed-point iteration
\begin{equation}
	\mathbf{V} \leftarrow (1-\alpha)\,\mathbf{V} + \alpha\big(V_s\mathbf{1} - \mathbf{M}\,(\mathbf{S}\oslash\mathbf{V})^{*}\big),
	\label{eq:fixedpoint}
\end{equation}
started from $\mathbf{V}=V_s\mathbf{1}$, where $\mathbf{M}$ is the calibrated shared-path matrix of \eqref{eq:Mmatrix} and $\oslash$ denotes elementwise division; a few iterations suffice, and a failure to converge, or any $|V_i|>V_s$, flags an infeasible state. The minimum section voltage is then $\min_i|V_i|$, and the section is admissible when this exceeds $U_{\min}$. The available power for train $i$, the $N$-train counterpart of the screen \eqref{eq:headroom}, is the largest $P_i$ for which every train voltage remains at or above $U_{\min}$ with the other demands held fixed; it is found by bisection on $P_i$ with \eqref{eq:sharedpath} re-solved at each trial. As with the screen, an already-infeasible state returns zero headroom, signalling that coordinated curtailment rather than a single-train margin is required. The complete evaluation is a small matrix assembly and a handful of iterations, with no admittance-matrix factorisation and a cost that scales with the number of trains rather than with the route's electrical complexity. Figure \ref{fig:flow} summarises the procedure.

The full AC power-flow model of Section \ref{sec:model}, exercised through the simulation framework, underwrites the estimate in two ways. A single-train sweep supplies the envelope $P_{\max}(d)$ from which $\beta(d)$ is calibrated, and a short set of two-train sweeps supplies the data from which $\gamma(d_i,d_k)$ is fitted; matched multi-train snapshots, identical train positions and powers presented to both the power flow and \eqref{eq:sharedpath}, are then used to confirm that the estimated per-train headrooms agree with those of the solver. The estimate is therefore a fast screen underwritten by, not a substitute for, the power flow: it captures the radial series-impedance behaviour that governs the minimum voltage, with the bonding-induced weakening of the inter-train coupling absorbed by $\gamma$, but it does not represent the current-limitation knee of \eqref{eq:acl} once voltages collapse, nor regenerative, photovoltaic or storage infeeds. The calibrated $\beta(d)$ and the pair $(\gamma_\infty,L)$ are stored together and are specific to a feeding configuration; a fresh offline calibration yields new values for a different switching or outage state. These are the bounds within which it is applied.

\subsection{Real-Time Use and Data Requirements}
In operation the estimate requires only two inputs per train: its distance from the supply point, available from position telemetry such as that already carried by connected driver advisory systems, and its instantaneous power, either measured or inferred from speed through \eqref{eq:te}--\eqref{eq:pdem}. The two calibration stages are performed once, offline, and do not enter the online path; each subsequent query avoids a network-wide power-flow solution. The two-train screen of \eqref{eq:headroom} reduces to a lookup in the precomputed envelope and a subtraction, and the full shared-path estimate to assembling a small $N\times N$ impedance matrix and iterating the fixed point of \eqref{eq:fixedpoint}, the work scaling with the number of trains $N$ rather than with the size or electrical complexity of the network. It thus replaces the repeated admittance-matrix factorisation of an online power flow with a fixed, network-independent amount of computation. This structural reduction is what makes the estimate a candidate for use within a control loop, where solving the full power flow at every step would not be practical; the resulting per-query cost, and its comparison against the solver, are examined in Section \ref{sec:cost}.

\section{Simulation Setup}\label{sec:setup}
The method is exercised on a representative corridor abstracted from GB mainline practice rather than any specific route.
A \SI{100}{\kilo\metre} section of double track is fed from a single \SI{25}{\kilo\volt} AC supply point at one end. A single-end feed representing a degraded or outage feeding state, the demanding configuration for available-power analysis, with paralleling posts along the route and a feeder-station capability of \SI{50}{\mega\watt} at the terminals. Rolling stock is represented by an electric multiple unit (EMU) whose parameters, taken from a single-train run over the route, are summarised in Table \ref{tab:stock}: a peak traction demand of \SI{4.44}{\mega\watt} (regenerating symmetrically), a \SI{0.4}{\mega\watt} auxiliary load, and a maximum speed of \SI{160}{\kilo\metre\per\hour}. The per-service demands used in the validation snapshots span auxiliary-dominated coasting through cruising at \SIrange{2}{3}{\mega\watt} to about \SI{5}{\mega\watt}, the upper values representing multiple-unit formations rather than a single unit. Available power is evaluated against the EN 50163 lowest non-permanent voltage $U_{\min}=\SI{17.5}{\kilo\volt}$ for \SI{25}{\kilo\volt} AC.

The calibration data are produced once for this configuration by the offline solver. A single-train sweep at \SI{1}{\kilo\metre} steps over the route yields the envelope $P_{\max}(d)$ from which $\beta(d)$ is recovered. Two background trains, placed at \SI{30}{\percent} and \SI{70}{\percent} of the route length and drawing \SI{5}{\mega\watt} and \SI{4}{\mega\watt} respectively, are each held fixed while a probe train is swept along the route; matching the estimate's two-train available power to the solver's over these sweeps fits the coupling factor $\gamma(d_i,d_k)$.

\begin{table}[!t]
	\centering
	\caption{Representative rolling-stock and feeder parameters.}
	\label{tab:stock}
	\begin{tabularx}{\columnwidth}{@{}lX r@{}}
		\toprule
		Symbol & Quantity & Value \\
		\midrule
		$m$ & Train mass (representative) & \SI{400}{\tonne} \\
		$P_{\mathrm{r}}$ & Peak traction power (EMU unit) & \SI{4.44}{\mega\watt} \\
		$P_{\mathrm{aux}}$ & Auxiliary load (EMU) & \SI{0.4}{\mega\watt} \\
		$F_{\max}$ & Maximum tractive effort & \SI{320}{\kilo\newton} \\
		$v_{\max}$ & Maximum speed & \SI{160}{\kilo\metre\per\hour} \\
		$A,B,C$ & Davis coefficients & representative \\
		$V_{s}$ & Supply (no-load) voltage & \SI{27.5}{\kilo\volt} \\
		$U_{\min}$ & Voltage-compliance limit (EN 50163) & \SI{17.5}{\kilo\volt} \\
		$\varphi$ & Displacement factor & \num{0.96} \\
		$S_{\mathrm{r}}$ & Feeder-station rating & \SI{50}{\mega\watt} \\
		\bottomrule
	\end{tabularx}
\end{table}

The demonstration uses a mixed-traffic scenario: a multi-train timetable in which several services run in both directions, creating frequent overlaps of acceleration and braking along the route, which drive the network into the low-voltage regime where the automatic current limitation engages.

\section{Results and Verification}\label{sec:results}

\subsection{Model Verification}
The model implements the EN 50388-1 automatic current limitation of \eqref{eq:acl}, under which a train's permissible current derates once its pantograph voltage falls below the upper threshold $V_1\approx\SI{19}{\kilo\volt}$. This behaviour is exercised by the mixed-traffic scenario examined below: the overlapping demands drive the minimum pantograph voltage to about \SI{15.6}{\kilo\volt} (Fig. \ref{fig:scenarioB}), well below the threshold at which limitation engages, yet the power flow remains bounded and convergent. It is the current limitation that produces this bounded behaviour, progressively reducing the current drawn as the voltage falls, where an unlimited constant-power load would instead drive the iteration to non-convergence and an unphysical collapse. The current-limitation model is therefore what distinguishes a credible low-voltage representation from a constant-power approximation; the latter, in which \eqref{eq:acl} is not enforced, is retained only for legacy protection-coordination studies and is not used here.

\subsection{Distance-Dependent Envelope}
Sweeping a single train's power demand to the voltage-compliance limit at each distance yields the envelope $P_{\max}(d)$ of Fig. \ref{fig:envelope}.
Near the supply point the deliverable power is limited by the feeder-station rating at \SI{50}{\mega\watt}; beyond a few kilometres the voltage limit binds and the deliverable power decays with distance, consistent with \eqref{eq:envelope}. This curve is the lookup table on which the online estimate operates.

\subsection{Envelope and Screening Under Mixed Traffic}
\begin{figure}[!t]
	\centering
	\includegraphics[width=\columnwidth]{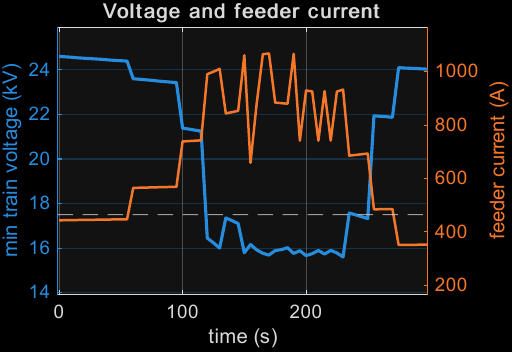}
	\caption{Mixed-traffic scenario: minimum train (pantograph) voltage and total feeder current over the run. The worst overlaps drive the minimum voltage below the \SI{17.5}{\kilo\volt} compliance limit (dashed) as the feeder current approaches \SI{1100}{\ampere}.}
	\label{fig:scenarioB}
\end{figure}

In the mixed-traffic scenario the overlapping acceleration of several services drives the minimum pantograph voltage down to about \SI{15.6}{\kilo\volt} during the worst overlaps, below the compliance limit $U_{\min}=\SI{17.5}{\kilo\volt}$ and well into the regime in which the automatic current limitation is active, while the feeder current peaks near \SI{1070}{\ampere} (Fig. \ref{fig:scenarioB}). The deepest voltage minima coincide with the intervals in which the largest demands fall furthest from the supply point, as the power-$\times$-distance argument of Section \ref{sec:analysis} predicts.

Applying the single-train envelope and the two-train screen of \eqref{eq:headroom} to the dominant interacting pair at each overlap identifies the instants at which the spare power at the reference location is exhausted. That is when the network cannot support the pair's combined demand at their locations. These instants coincide with the overlaps responsible for the voltage minima, confirming that the envelope and screen localise the binding events correctly. The limitations of the screen for simultaneous interactions among three or more trains are as set out in Section \ref{sec:method}.

\subsection{Comparison of the Headroom Estimators}\label{sec:comparison}
The estimators of Section \ref{sec:method} return different quantities at different cost, as summarised in Table \ref{tab:methods}. The power-flow snapshot is the reference: holding all other demands fixed, it returns each train's exact available power by bisection, but at the cost of a power-flow solution at every step. The single-train envelope and the two-train screen are the cheapest, yet address only one train or the dominant interacting pair. The shared-path estimate closely approximates the per-train available power of the snapshot. In addition, it returns the minimum section voltage for any number of trains and with no power-flow solution.

\begin{table*}[!t]
	\centering
	\caption{What each headroom estimator computes. The envelope and the two-train screen are read from the stored $P_{\max}(d)$ lookup; the shared-path estimate and the power-flow snapshot operate on the live train state.}
	\label{tab:methods}
	\footnotesize
	\begin{tabularx}{\textwidth}{@{}X X l X X@{}}
		\toprule
		Estimator & Returns & Inputs & Multi-train fidelity & Online cost \\
		\midrule
		Power-flow snapshot (reference) & Per-train available power, all other demands fixed & $d_i,\,P_i$ (all trains) & Exact ($N$-train) & High: a power-flow solution per bisection step \\
		Single-train envelope $P_{\max}(d)$ & Maximum power one isolated train may draw, versus distance & $d$ & Single train only & Lookup, computed once offline \\
		Two-train screen, \eqref{eq:headroom} & Spare power at a reference location for a following train & $d_A,\,P(d_A)$ & Dominant pair only (conservative) & Lookup and subtraction \\
		Shared-path estimate & Per-train available power and minimum section voltage & $d_i,\,P_i$ (all trains) & Approximate ($N$-train); mean $|\Delta|\approx\SI{9}{\percent}$, max $\approx\SI{16.6}{\percent}$ here & Matrix build and a few iterations; scales with $N$, not network \\
		\bottomrule
	\end{tabularx}
\end{table*}

To quantify the accuracy of the shared-path estimate it is compared against the power-flow snapshot on matched multi-train cases: identical train positions and powers are presented to both, and the per-train available power $H_i$ is recorded from each. This is an approximation-accuracy check against the same power-flow model, establishing the estimate's internal consistency with the solver it is calibrated to, rather than validation against independent or measured data. Table \ref{tab:deviation} reports the reference value, the shared-path value, and the deviation
\begin{equation}
	\Delta_i = \frac{H_i^{\text{shared}} - H_i^{\text{ref}}}{H_i^{\text{ref}}}\times 100\,\%,
	\label{eq:deviation}
\end{equation}
in which a positive value indicates an optimistic estimate. By construction the calibrated single-train envelope matches the power-flow envelope to numerical tolerance (Section \ref{sec:multitrain}). Because $\beta$  is fixed exactly by the envelope, these multi-train deviations probe only the two-parameter coupling model $(\gamma_\infty, L)$; reproducing sixteen per-train values across two-, three-, and four-train snapshots from a two-parameter fit is therefore a test of generalisation rather than of fit. For the single-end-fed corridor the mutual fit returns $\gamma_\infty\approx\num{0.02}$ and $L\approx\SI{3.4}{\kilo\metre}$: the coupling collapses within a few kilometres of separation, so trains more than roughly \SI{10}{\kilo\metre} apart are nearly independent and each binds on its own voltage, which is what the solver shows, and what the uncorrected $\gamma\equiv1$ model failed to capture.

Five operating snapshots are used, spanning two-, three-, and four-train states from off-peak to peak loading and from a near-substation cluster to a route-wide spread (Table \ref{tab:deviation}). Across all sixteen per-train values the estimate tracks the reference to a mean absolute deviation of about \SI{9}{\percent} and a maximum of \SI{16.6}{\percent}, against deviations of tens of per cent for the same model before the mutual calibration. Two features are worth noting. First, the accuracy improves as more trains share the feeder, the four-train cases are the closest, because the binding constraint is distributed across trains and the per-train errors partly cancel, the favourable direction for a realistic busy section. Second, the residual error is no longer one-signed: the heavily loaded near-substation train, where the largest currents make the single-scalar diagonal calibration least exact, remains conservative (deviations of $-11$ to $-17\,\%$), whereas distant lightly loaded trains in the route-wide spread turn modestly optimistic ($+7$ to $+11\,\%$). The estimate is therefore a tight two-sided approximation rather than a strictly conservative bound; where a guaranteed-safe margin is required, for example as a hard constraint in the demand-shaping optimisation, the two-train screen of \eqref{eq:headroom}, which lies below both, is retained for that role as set out in Section \ref{sec:method}.

\begin{table}[!t]
	\centering
	\caption{Per-train available power for matched multi-train snapshots on the \SI{100}{\kilo\metre} single-end-fed corridor: power-flow reference (PF) against the calibrated shared-path estimate, with the deviation \eqref{eq:deviation} (positive $=$ optimistic). Mean $|\Delta|\approx\SI{9}{\percent}$, maximum \SI{16.6}{\percent}.}
	\label{tab:deviation}
	\footnotesize
	\begin{tabular}{@{}llrrr@{}}
		\toprule
		Snapshot & Train ($d$, $P$) & PF & Shared & $\Delta$ \\
		         &                  & (MW) & (MW) & (\%) \\
		\midrule
		S1 off-peak      & \SI{20}{\kilo\metre}, \SI{3.5}{\mega\watt} & 26.0 & 22.0 & $-15.3$ \\
		(2 trains)       & \SI{60}{\kilo\metre}, \SI{2.0}{\mega\watt} & 10.9 &  9.1 & $-16.6$ \\
		\midrule
		S2 inter-peak    & \SI{10}{\kilo\metre}, \SI{4.5}{\mega\watt} & 28.7 & 24.5 & $-14.7$ \\
		(3 trains)       & \SI{40}{\kilo\metre}, \SI{3.0}{\mega\watt} & 14.1 & 12.9 &  $-8.3$ \\
		                 & \SI{75}{\kilo\metre}, \SI{1.5}{\mega\watt} &  7.5 &  7.0 &  $-7.2$ \\
		\midrule
		S3 peak          & \SI{8}{\kilo\metre},  \SI{5.0}{\mega\watt} & 30.9 & 26.8 & $-13.1$ \\
		(3 trains)       & \SI{30}{\kilo\metre}, \SI{3.5}{\mega\watt} & 14.8 & 13.7 &  $-7.0$ \\
		                 & \SI{60}{\kilo\metre}, \SI{2.5}{\mega\watt} &  9.9 &  9.1 &  $-7.8$ \\
		\midrule
		S4 peak spread   & \SI{12}{\kilo\metre}, \SI{4.0}{\mega\watt} & 24.0 & 22.9 &  $-4.8$ \\
		(4 trains)       & \SI{38}{\kilo\metre}, \SI{3.0}{\mega\watt} & 11.7 & 12.6 &  $+7.3$ \\
		                 & \SI{62}{\kilo\metre}, \SI{2.2}{\mega\watt} &  7.8 &  8.5 &  $+8.6$ \\
		                 & \SI{88}{\kilo\metre}, \SI{1.2}{\mega\watt} &  5.4 &  6.0 & $+11.3$ \\
		\midrule
		S5 dense         & \SI{6}{\kilo\metre},  \SI{4.0}{\mega\watt} & 34.2 & 30.4 & $-11.2$ \\
		near-sub         & \SI{16}{\kilo\metre}, \SI{3.5}{\mega\watt} & 22.4 & 21.0 &  $-5.9$ \\
		(4 trains)       & \SI{28}{\kilo\metre}, \SI{2.8}{\mega\watt} & 15.1 & 14.4 &  $-5.1$ \\
		                 & \SI{45}{\kilo\metre}, \SI{2.0}{\mega\watt} & 11.4 & 10.8 &  $-5.2$ \\
		\bottomrule
	\end{tabular}
\end{table}

\subsection{Computational Behaviour}\label{sec:cost}
\begin{figure}[!t]
	\centering
	\includegraphics[width=\columnwidth]{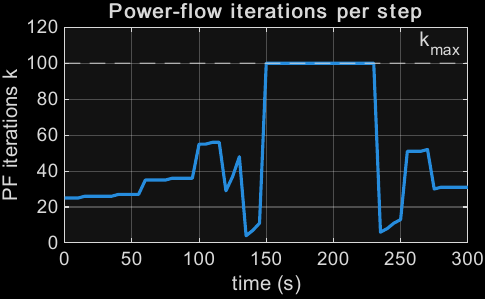}
	\caption{Mixed-traffic scenario: power-flow iterations per time step. Convergence is rapid under light load but rises to the cap $k_{\max}$ during the most stressed overlaps, the numerical signature of the voltage stress the estimate is designed to anticipate cheaply.}
	\label{fig:iters}
\end{figure}

The cost of anticipating these events by power flow is shown by the solver's behaviour: the number of iterations required for convergence rises sharply during the high-load, low-voltage overlaps and reaches the preset maximum $k_{\max}$ in the most stressed intervals, here in 17 of the 61 time steps (Fig. \ref{fig:iters}).

By contrast the online estimate evaluates either the screen of \eqref{eq:headroom} as a lookup and a subtraction, or the shared-path model of \eqref{eq:sharedpath} as a small matrix assembly and a handful of fixed-point iterations, at a cost that scales with the number of trains and not with the electrical complexity of the network.

The two regimes differ structurally. The full power flow assembles and factorises an admittance matrix whose order is set by the number of electrical nodes $M$ used to represent the feeder: supply points, paralleling and autotransformer posts, and conductor segments. A new solution is required at every trial of the bisection that locates a train's available power, so its cost grows with the size and electrical detail of the network irrespective of how many trains are present. The estimate carries no such matrix: one query assembles the $N\times N$ shared-path matrix of \eqref{eq:Mmatrix}, with $N$ the number of trains in the section (typically a handful), and iterates the damped fixed point of \eqref{eq:fixedpoint} to convergence, each iteration costing $\mathcal{O}(N^2)$ with an iteration count that stays small and largely independent of $N$. A single minimum-voltage feasibility evaluation is therefore $\mathcal{O}(N^2)$, and the complete set of per-train available powers adds one bounded bisection per train, giving $\mathcal{O}(N^3)$ polynomial in the handful of trains and, decisively, independent of $M$.

Table \ref{tab:timing} reports measured per-query execution times for the network of Section \ref{sec:setup}, separating the minimum-voltage feasibility query that a control loop would call most often from the heavier full per-train set, alongside the same per-train computation driven by the power flow. The feasibility query completes in well under \SI{0.1}{\milli\second} at every $N$ tested, and the full per-train set is obtained roughly an order of magnitude faster than the power-flow reference. Over the handful of trains in a feeding section the estimate's per-train cost grows far more gently than the cubic worst case, being dominated by the bisection performed once per train rather than by the matrix iteration. The estimate's demand is therefore bounded and predictable where that of repeated power-flow solution is neither, which is what makes it a candidate for evaluation within a control loop.

\begin{table}[!t]
	\centering
	\caption{Measured per-query execution time of the solver-free estimate against the full power flow computing the same quantity on the same snapshot, versus the number of trains $N$, on one machine. The minimum-voltage query is a single feasibility evaluation; the per-train sets return the available power of every train.}
	\label{tab:timing}
	\footnotesize
	\setlength{\tabcolsep}{4pt}
	\begin{tabular}{@{}rrrrr@{}}
		\toprule
		    & Min-voltage & \multicolumn{2}{c}{Per-train set (\si{\milli\second})} & Speed-up \\
		\cmidrule(lr){3-4}
		$N$ & query (\si{\milli\second}) & Estimate & Power flow & ($\times$) \\
		\midrule
		1  & 0.016 & 1.9   & 208  & 110 \\
		2  & 0.062 & 19.1  & 322  & 17  \\
		4  & 0.045 & 45.8  & 573  & 13  \\
		8  & 0.053 & 108.5 & 1272 & 12  \\
		\bottomrule
	\end{tabular}
\end{table}

\section{Discussion}\label{sec:discussion}
The method buys real-time observability of available power, with limitations that bound its present scope. The envelope $P_{\max}(d)$ is precomputed and therefore assumes a fixed feeding configuration; switching states or supply outages would require recomputation or a small family of envelopes selected by configuration. More fundamentally, the envelope is exact only for a single train: as shown in Section \ref{sec:method}, available power does not compose additively for multiple trains, because the voltage at each train depends on the positions as well as the powers of the others, and the two-train screen of \eqref{eq:headroom} is retained only as a conservative proxy for the dominant pairwise interaction rather than a general $N$-train headroom. This restriction is lifted by the solver-free multi-train estimate of Section \ref{sec:multitrain}, whose shared-path voltage model reproduces the single-train envelope exactly by construction and, after a second calibration stage that fits the inter-train coupling from a handful of two-train solver sweeps, tracks the full power flow to a mean absolute deviation of about \SI{9}{\percent} across realistic multi-train snapshots. Its remaining approximations bound its scope: the calibration is per-configuration; the separation-dependent coupling factor is a smooth average of a pattern that in truth varies with the detailed geometry, so the estimate is a tight two-sided approximation rather than a strictly conservative bound, slightly optimistic for distant lightly loaded trains; and the current-limitation knee and regenerative or stored-energy infeeds are omitted. These motivate continued validation against the solver rather than its wholesale replacement, and the retention of the strictly conservative two-train screen wherever a guaranteed-safe margin is required. The representative corridor and single-feeder topology are deliberate abstractions; the envelope formulation extends to autotransformer feeding through the equivalent network, and to multiple supply points by sectionalising the envelope. Where the residual two-sided error is unacceptable, the coupling term could instead be taken directly from a reduced nodal impedance matrix of the solved network, which is exact for a linear feeder; the present separation-dependent factor is preferred here as the cheaper calibration that needs only two-train sweeps. The estimate's accuracy ultimately depends on the fidelity of the underlying power-flow model and on the availability of timely position and power data, the latter being well matched to connected driver advisory systems.

\section{Conclusion}\label{sec:conclusion}
This paper has made available traction power observable in real time on a mixed-traffic AC railway. An AC power-flow model incorporating the EN 50388-1 current-limitation characteristic was used to establish that minimum network voltage is governed by power and distance jointly, and from it a rigorous single-train available-power envelope was derived. We further showed that this envelope does not compose additively for multiple trains, and explained the coupling responsible, retaining a conservative two-train screen for the dominant interaction. The model's low-voltage current-limitation behaviour was verified against the EN 50388-1 characteristic, and the envelope and screen were demonstrated on a representative corridor without recourse to repeated power-flow solution. Because the estimate is location-aware and its online cost scales with the number of trains rather than with the network, it provides a practical input for grid-coupled control. The single-train envelope was further generalised to a solver-free multi-train estimate through a calibrated shared-path voltage model. Its calibration is in two offline stages: a single-train sweep that fixes the self-impedance and reproduces the envelope, and a short set of two-train sweeps that fixes the inter-train coupling the first stage cannot observe, after which it returns the per-train available power for any number of trains without an online solution, within a mean of about nine per cent of the full power flow and improving as more trains share the feeder. This estimate provides the foundation for a demand-shaping control layer that embeds the power-flow model within an optimisation, shaping multi-train demand through a driver advisory system subject to timetable constraints.

\printbibliography

\end{document}